\documentclass[twocolumn,prl,showpacs,preprintnumbers,amsmath,amssymb]{revtex4}
\usepackage{graphicx}% Include figure files
\usepackage{dcolumn}% Align table columns on decimal point
\usepackage{bm}% bold math
\usepackage{longtable}
\usepackage{epsfig}

\begin{document}

\title{Synthesis of quantum-confined CdS nanotubes}

\author{A. K. Mahapatra\footnote{e-mail: amulya@iopb.res.in}}
\affiliation{Institute of Physics, Sachivalaya Marg, 
Bhubaneswar, 751005, India}

\begin{abstract}
CdS nanotubes with wall thickness comparable to excitonic diameter of 
the bulk  material are synthesized by a chemical route. A change in 
experimental conditions result in formation of nanowires, and well-separated 
nanoparticles. The diameter and wall thickness of nanotubes measured to be 
14.4 $\pm$ 6.1 and 4.7 $\pm$ 2.2 nm, respectively. A large number of CdS 
nanocrystallites having wurzite structure constitute these nanotubes. These 
nanotubes show  
high energy shifting of optical absorption and  photoluminescence peak 
positions, compared to its bulk value, 
due to quantum confinement effect. It is proposed that nucleation and growth of
bubbles and particles in the chemical reaction, and their kinetics and 
interactions are responsible for the formation of nanotubes. 
\end{abstract}


\maketitle
\section*{Introduction}
	The discovery of carbon nanotubes (Iijima 1991) has generated 
considerable research interest to synthesize such type of tubular 
nanostructures of other materials and study its properties. Carbon 
nanotubes are conceptualized as the wrapping of graphite layers into a 
seamless cylinder. Synthesis of single crystalline nanotubes of other similar 
type of layered materials like BN (Chopra et al. 1995), $MoS_2$ (Feldman et al. 1995) and $WS_2$(Tenne et al. 1992) are reported. In these materials, there
exists a strong force within the layer plane, but a weak van der waals force 
between the inter-layer planes. It helps such materials to self assemble 
in the form of nanotubes. However, comparatively more isotropic materials 
like CdS (Zhan et al. 2000), CdSe (Duan and Lieber 2000), 
GaAs (Duan et al. 2000), and Si (Yu et al. 1998) 
tend to form nanowires instead of nanotubes. Therefore,
nanotubes of these isotropic materials are generally synthesized by using 
nanowires as templates. CdS nanotubes are also synthesized by using Sn nanowires as the 
template (Hu et al. 2005). However, those synthesized CdS nanotubes have a wall 
thickness much larger than the excitonic diameter of the bulk CdS (6 nm).
 Hence, no quantum confinement effects could  be observed. Aspect ratio and 
wall thickness of nanotubes play a major role in its mesoscopic properties
(Masale et al. 1992). The quantum confinement effect can be observed 
if the wall thickness of nanotube is comparable to the excitonic diameter of 
bulk material. It should be noted that quantum confinement effects are least 
studied in one-dimensional systems, particularly with tubular structure,
 as compared to other low-dimensional systems.

	 The present work reports a single-step chemical process to synthesize
micron length CdS nanotubes with  wall thickness comparable to the  excitonic 
diameter of the bulk material.  A change in experimental conditions result in 
formation of nanowires, and well separated nanoparticles. A mechanism of 
formation of the nanotube is described. It is proposed that
bubbles are playing the key role in formation of nanotubes. Optical absorption 
and photoluminescence measurements are carried out in order to study the 
 quantum confinement effect. 

\section*{Experimental}
The precursors used for the synthesis of CdS nanotubes  are thiourea 
(NH$_2$CSNH$_2$), Cadmium sulfate (CdSO$_4$), ammonia (NH$_3$), and poly vinyl 
alcohol (PVA). 
0.01 M aqueous solution of CdSO$_4$ and thiourea are prepared separately. About 
5 mL of 3 M aqueous ammonia solution is added through a buret into 20 mL of 
CdSO$_4$ solution in a slow stirring condition. During addition of ammonia it 
turns to a turbid white solution and then becomes completely transparent. 
A total of 15 mL of 4$\%$  aquous 
solution of PVA (degree of polymerization: 1,700 - 1,800) is added to it.
 About 20 mL of thiourea solution is then put  in the same stirring condition, 
and left in the ambient condition without any further stirring. After around
10 min the solution becomes pale yellow color, which suggests formation of 
CdS. After 90 min, carbon coated Cu-grids are dipped into this 
yellow color solution and held aloft to dry. These dried grids 
are used for transmission electron microscopy (TEM) analysis. For other 
experiments nanotubes are collected by centrifugation.

  TEM is performed using JEOL-2010 operated at 200 KeV electron beam energy. X-ray diffraction (XRD) 
was conducted on a philips PW1877 diffractometer using Cu $K_\alpha$ radiation. Optical 
absorption measurements are carried out by using a dual beam Shimadzu 
UV-3101 PC spectrophotometer. The photoluminescence (PL) measurements are
 carried out using 369 nm line from a Oriel Hg/Xe lamp as the source 
of excitation. Luminescence is detected by a PMT detector attached with 
Jovin Yvon TRIAX-180 spectrometer. Experiments are carried out in ambient temperature.

 \section*{Results and discussions}  

\begin{figure}
\begin{center}
\includegraphics[width=7.0cm]{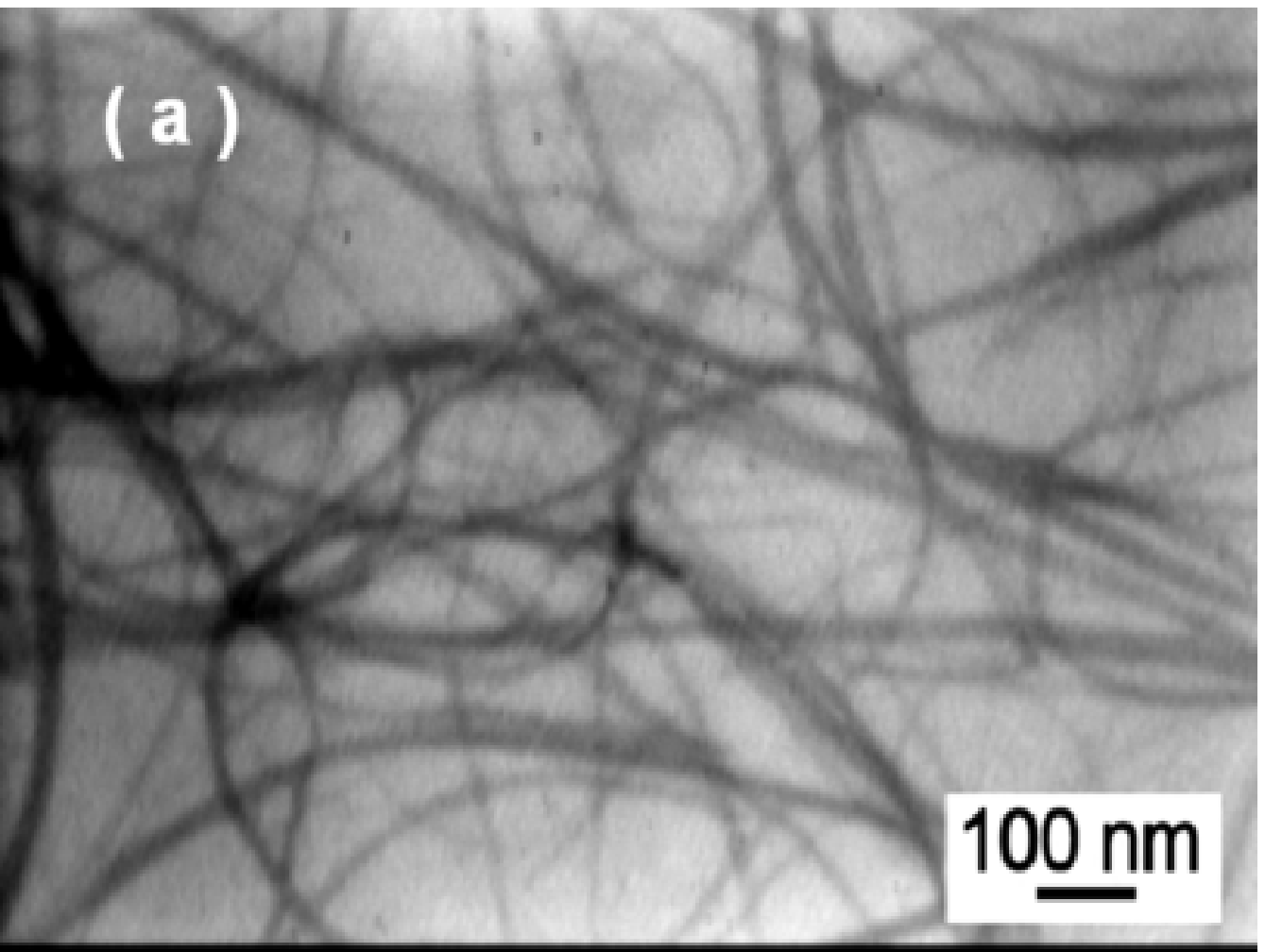}
\includegraphics[width=7.0cm]{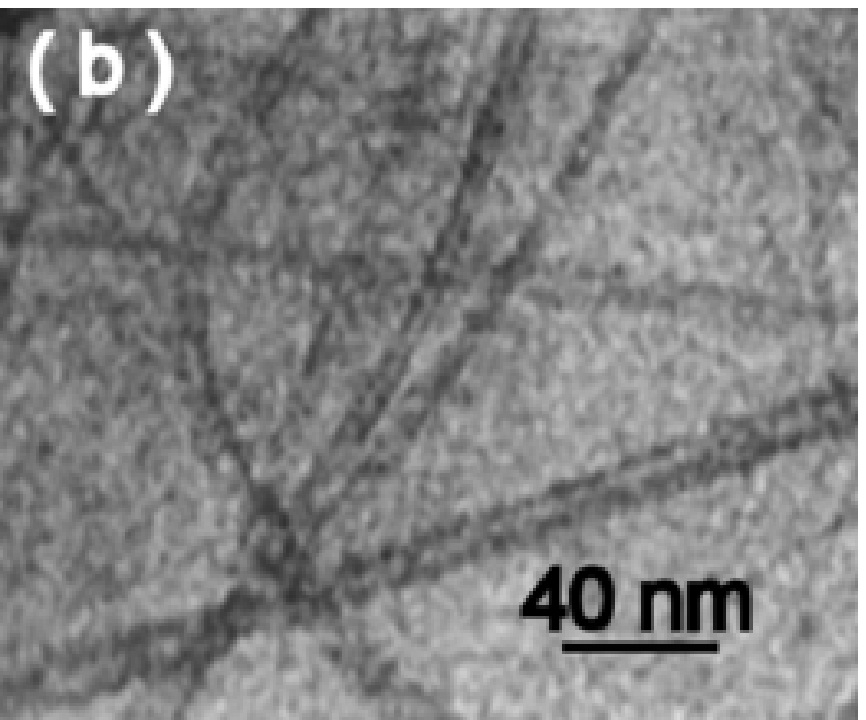}
\includegraphics[width=4.3cm]{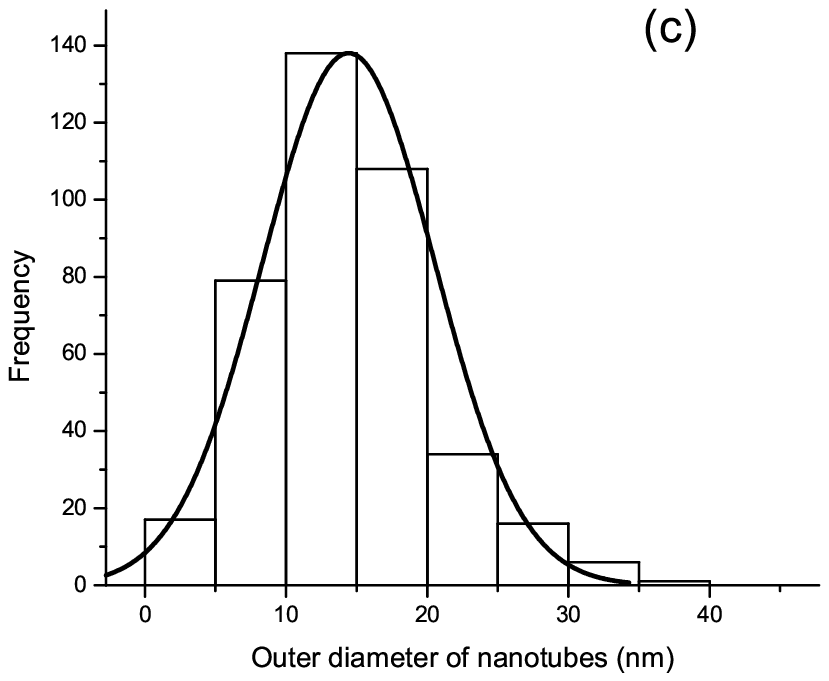}%
\includegraphics[width=4.3cm]{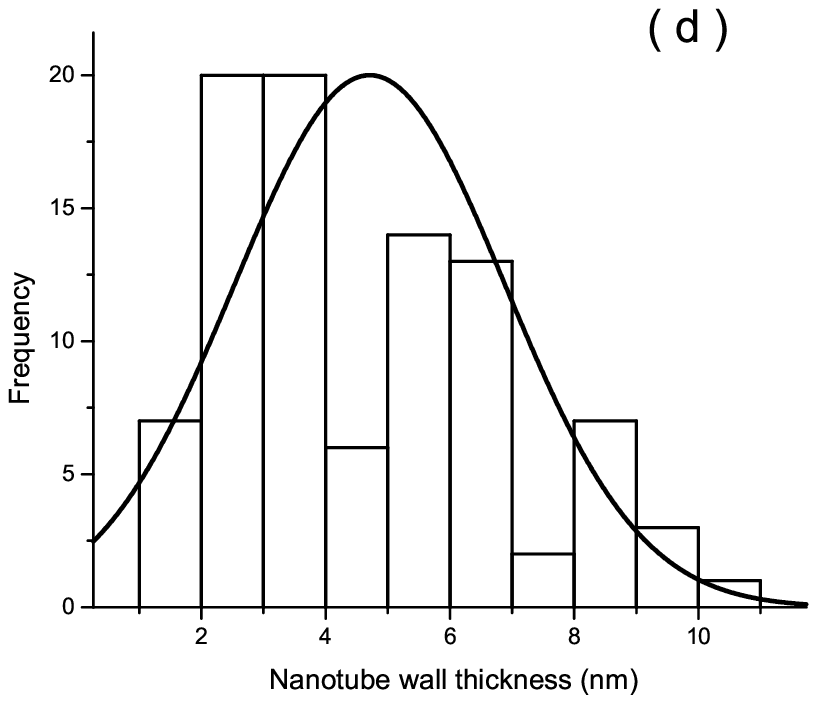}
\caption{TEM micro-graphs of  CdS nanotubes (a) with lower magnification 
(b) with higher magnification. (c) Histogram of outer diameter of nanotubes 
(d) histogram of thickness of nanotubes.}
\label{tem1}
\end{center}
\end{figure}

\begin{figure}
%\begin{center}
%\vspace*{-3cm}
\includegraphics[width=1.8cm]{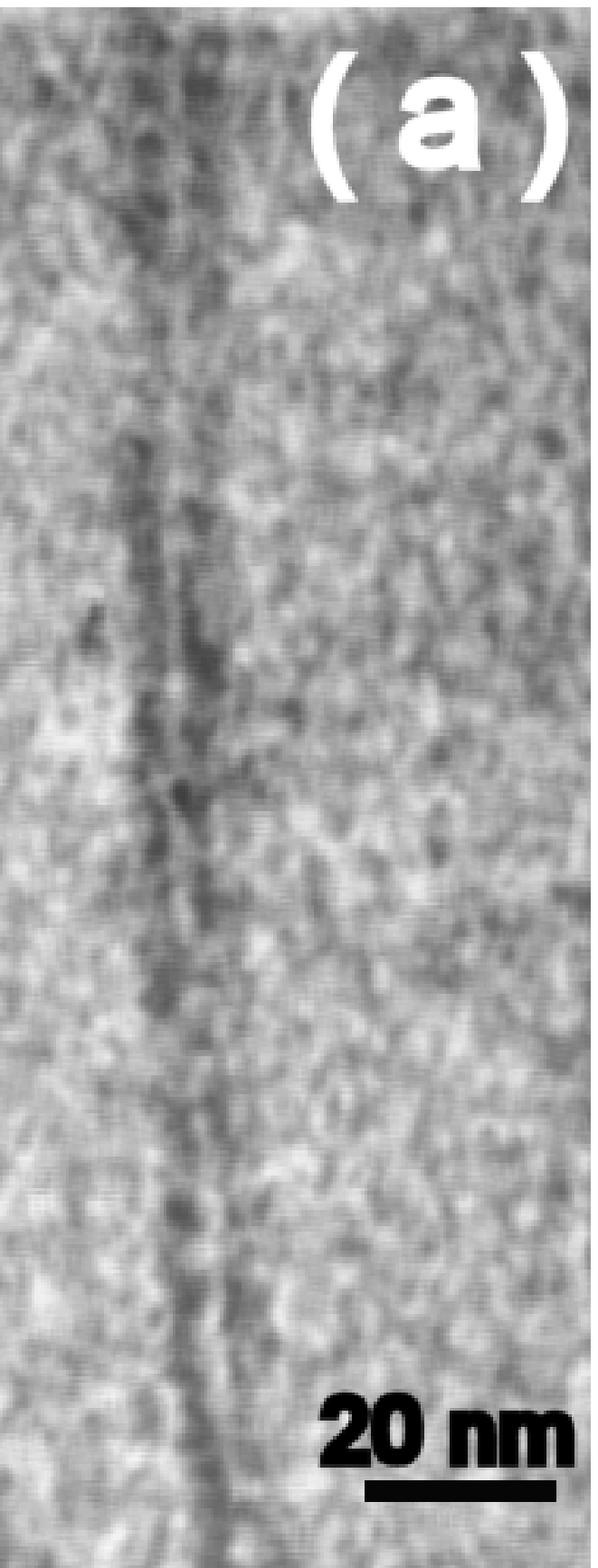}%
\hspace{0.1cm}
\includegraphics[width=6.4cm]{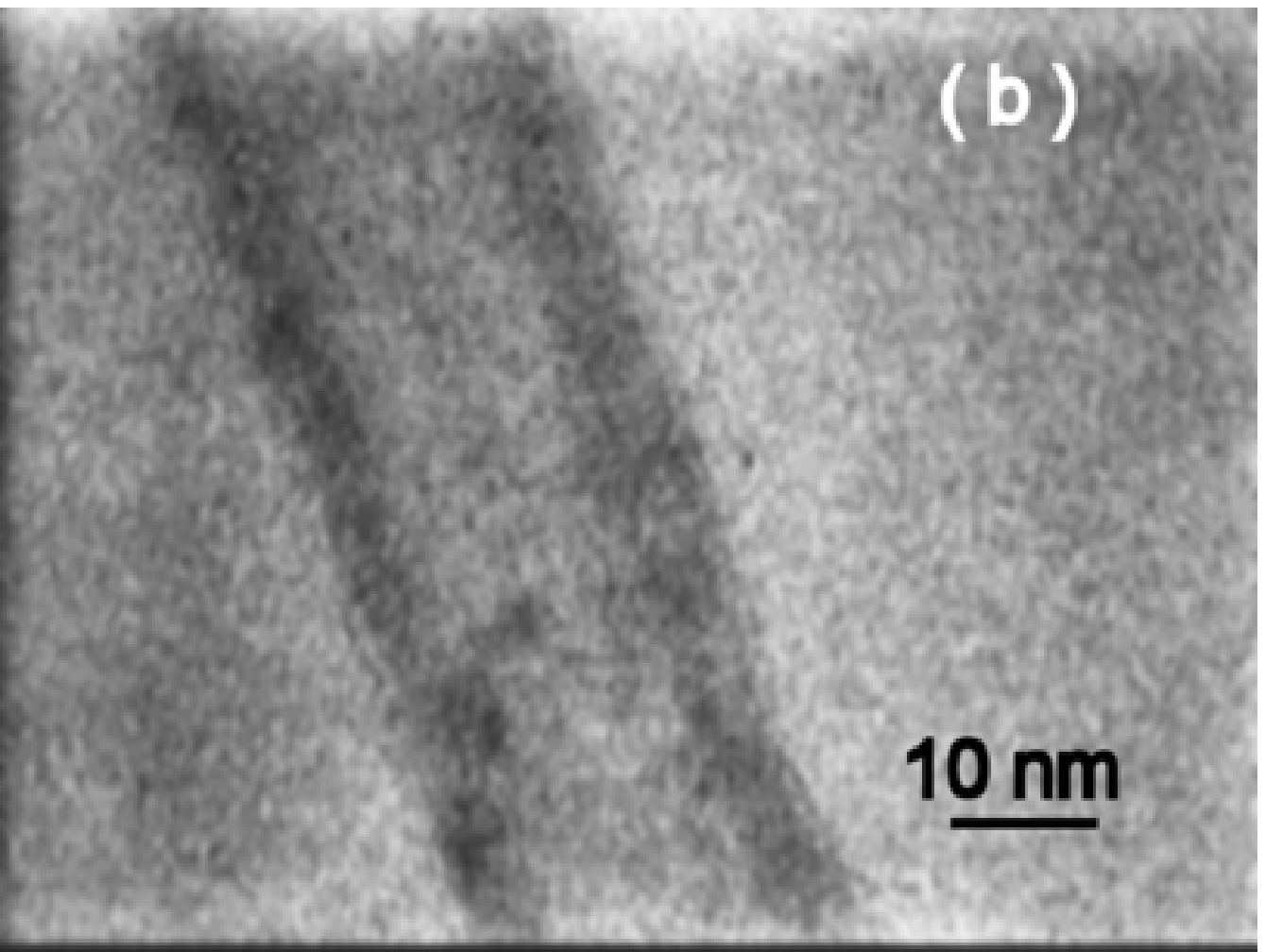}
\caption{TEM micro-graphs of (a) a  nanotube with smaller diameter 
(b) a nanotube with bigger diameter.}
\label{tem2}
%\end{center}
\end{figure}

The formation of CdS nanotubes is confirmed by transmission electron 
microscopy (TEM). The typical TEM micrographs at a lower and at a higher 
magnification are shown in Fig.\ref{tem1}a, b, respectively. From Fig.\ref{tem1}a, 
it is not possible to conclude whether these are 
nanotubes or nanowires. However, it is conformed that these nanostructures are 
of micron lengths and have  high aspect ratio. With 
higher magnification (Fig.\ref{tem1}b),  a  contrast between the solid 
side wall (darker contrast) and hollow middle part (lighter contrast) of these 
one-dimensional structure is observed. It gives the signature of formation of
nanotubes. Outer diameter and wall thickness of several nanotubes are measured.
A histogram of outer diameter
of CdS nanotubes along with a fitted Gaussian , is shown in Fig. \ref{tem1}c. 
The mean diameter of the nanotube is 14.4 nm with a standard deviation of 
6.1 nm. The histogram of wall thickness, along with a fitted Gaussian, is 
shown in Fig. \ref{tem1}d. The mean wall thickness of the nanotubes is 4.7 nm
with a standard deviation of 2.2 nm.  TEM image of a nanotube with a smaller 
diameter and a nanotube with a larger diameter are shown in Fig. \ref{tem2} a,b respectively. The fluctuation observed in the thickness 
of  wall is comparatively less than the fluctuation in  diameter of the 
nanotubes.

\begin{figure}
\begin{center}
\epsfxsize=8.8 true cm{\epsfbox{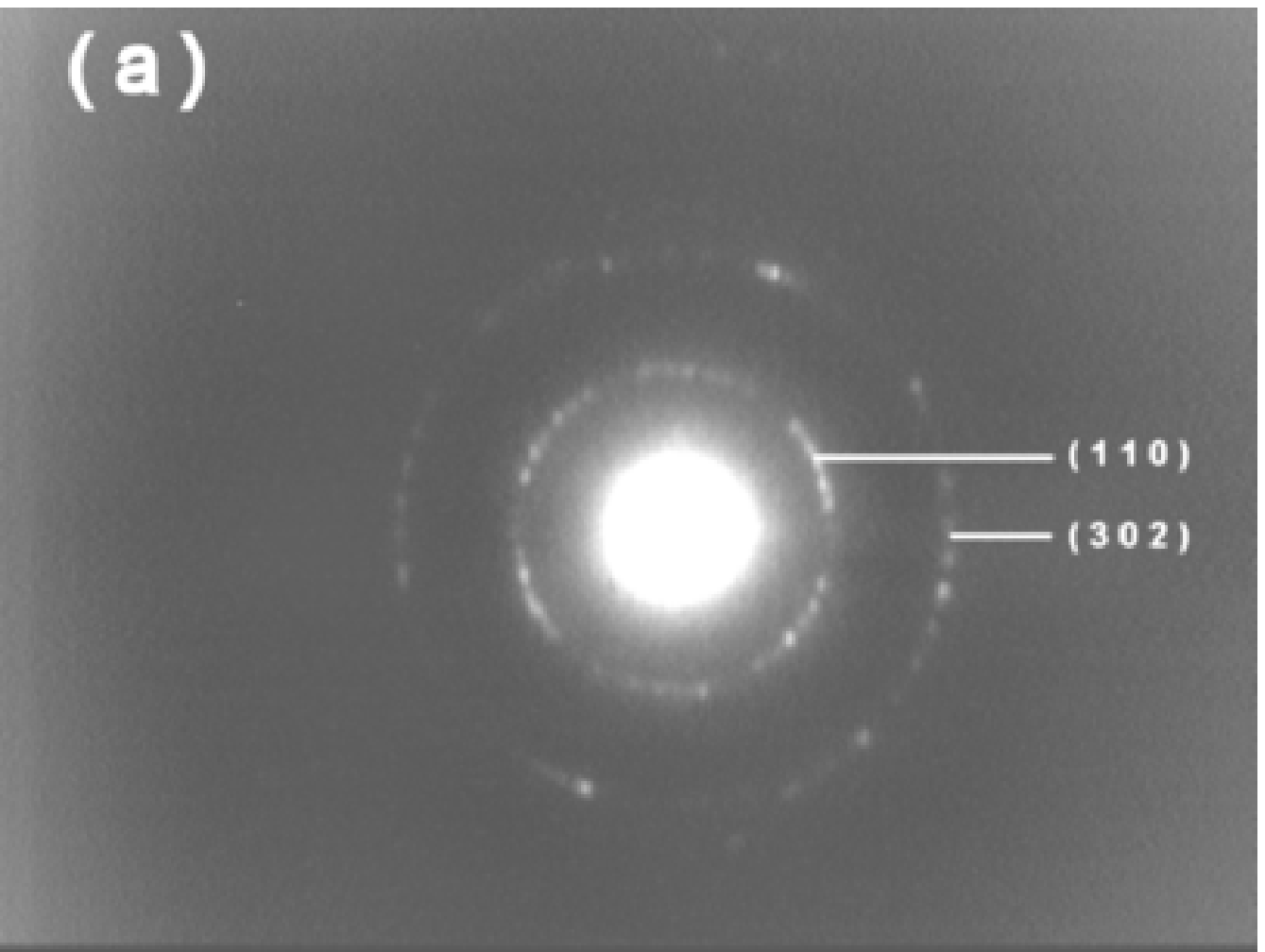}}
\epsfxsize=8.8 true cm{\epsfbox{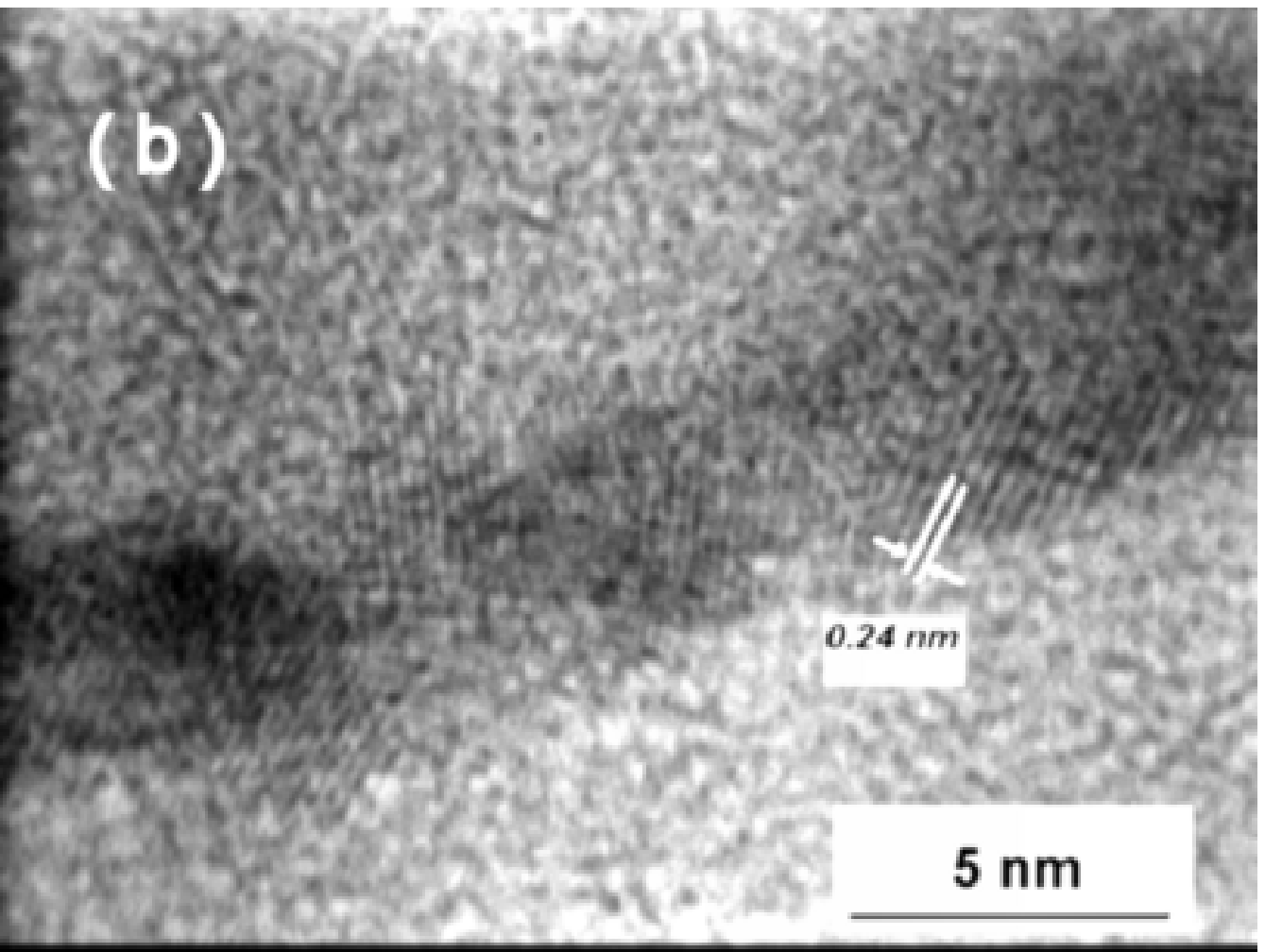}}
\caption{(a)Diffraction pattern of CdS nanotubes (b) HRTEM micro-graph of a 
portion of side wall in CdS nanotube.}
\label{tem3}
\end{center}
\end{figure}

\begin{figure}
\begin{center}
\epsfxsize=8.8true cm{\epsfbox{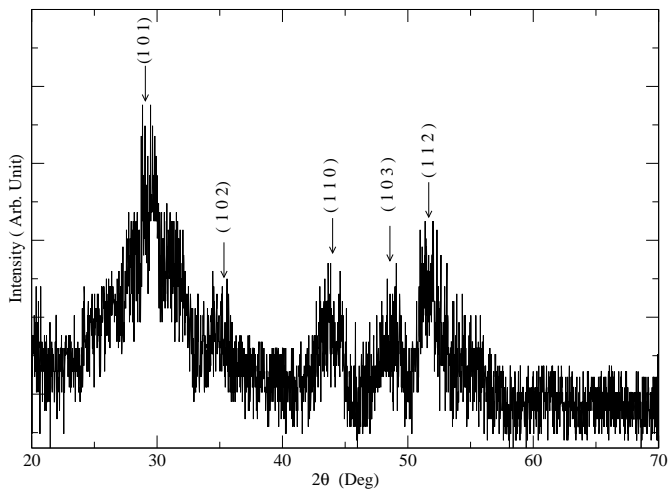}}
\caption{XRD spectrum of CdS nanotubes.}
\label{xrd}
\end{center}
\end{figure}

The electron diffraction (ED) pattern is shown in Fig. \ref{tem3}a.    
 The rings in ED pattern suggests polycrystalline nature of the nanotubes. 
These nanotubes are not single crystalline, but consists of numerous 
nanocrystallites  as it can be seen in high resolution electron micrograph 
(Fig.\ref{tem3}b). It seems large number of nanocrystallites bind together 
and form nanotube. X-ray diffraction (Fig. \ref{xrd})  peaks 
obtained are also broad 
due to small dimensions of these constituent crystallites. 
The XRD pattern shows characteristic peaks of wurzite structure of CdS 
(JCPDS card No. 41-1049). The 
interplanar distance of 0.24 nm, as shown in Fig.\ref{tem3}b, matches for 
(102)
planes and rings in ED pattern corresponds to (110) and (302) planes of 
wurzite structure of CdS. The formula used to calculate $d$-value in the ED 
pattern is $d_{hkl}$ =$\lambda L$/${R}$. where L is the Camera length, 
R is the radius of the ring and $\lambda$ is the wavelength of the electron 
beam.

The general chemical equation for the formation of CdS nanotube can be written 
as follows (Pavaskar et al. 1977; Hariskos et al. 2001)    :\\
\begin{math}
CdSO_4+4(NH_3)  \rightarrow  [Cd(NH_3)_4]SO_4  \\  
C(NH_2)_2S+ OH^- \rightarrow  CH_2N_2+H_2O+HS^{-}  \\
CH_2N_2+H_2O \rightarrow (NH_2)_2CO\\
(NH_2)_2CO + 2 OH^- \rightarrow CO_3^{2-} + 2NH_3(g)\\
HS^{-}+OH^-  \rightarrow  S^{--}+H_2O  \\
$~$ [Cd(NH_3)_4]^{++} + S^{--} \rightarrow  CdS (s)+4NH_3 (g) \\
\end{math}

	The chemical reaction mentioned above, leads to supersaturation of CdS 
concentrations. According to classical nucleation theory (Markov 1995),
the energy required 
to form a critical particle homogeneously is given by:
\begin{equation}
 W_p=\frac{1}{3}\sum_n\sigma_n A_n
\end{equation} 
where $A_n$ is surface area and $\sigma_n$ is specific surface free energy of 
the $n$-th surface of the particle. Since crystal faces with different 
crystallographic orientation have different specific surface energy, the 
crystallite form a shape in which the total surface energy will be minimum.

  It should be noted that the chemical route adopted to synthesize CdS 
nanotube is similar to 
the chemical bath deposition technique to make bulk CdS thin 
films (Pavaskar et al. 1977). The optimization parameters like concentration of
the reactant and working temperature are varied. PVA is also used during these
 chemical synthesis process. PVA is a surface active and water soluble polymer. 
Although, it does not take part in the chemical reaction, its addition 
increases viscosity (Briscoe et al. 2000)  and decreases surface tension 
(Bhattacharya and Ray 2004) of the solution. 

According to the Henry's law, the amount of a gas dissolved in a given volume of liquid,$C_{eq}$, is
 directly proportional to the partial pressure of that 
gas,$P$,
 above the liquid. It can be written as:
\begin{equation}
C_{eq}=kP
\end{equation} 

where $k$ is the Henry's law constant.

  As viscosity of the solution is increased due to addition of PVA, the 
diffusional escape of the ammonia gas decreased and higher concentration of 
 gas generated inside the solution than its equilibrium 
concentration. It results in supersaturation and nucleation of ammonia gas 
bubble inside the solution.

The energy required to form a critical
bubble homogeneously (Landau and Lifshitz 1999; Bowers et al. 1995) is given by
\begin{equation} 
 W_b= \frac{1}{3}\sigma A= \frac{16\pi{\sigma}^3 k^2}{3(C_s-C_{eq})^2}
\end{equation}
here $\sigma$ is the surface tension and $C_s$ is the supersaturation 
concentration.

 The probability of formation of critical bubble is 
proportional to exp$(-W_b/T)$. Due to presence of $\sigma^3$ term in the 
numerator of $W_b$ , the probability of formation of gas bubbles is 
increased significantly by the reduction of surface tension. 

   Although we considered the case of a homogeneous nucleation, the colloids 
( both bubbles and particles ) most likely nucleate heterogeneously, as later 
process is energetically favorable than the former. 

 These colloids  experience the 
gravitational forces due to its mass, the buoyant force due to the displaced
fluid and the frictional force due to viscosity of the medium. This leads to a  
terminal velocity at which these colloids will move and is given 
by the equation (Lamb 1945)
\begin{equation}
  V=\frac{2}{3}\cdot \frac{gr^2(\rho'-\rho)}{\eta}\cdot\frac{\eta+\eta'}{2\eta+3\eta'}
\end{equation}
 Here $\rho'$ and $\eta'$ are the density and dynamic viscosity 
      of the colloids, respectively. Similarly
      $\rho$ and $\eta$ are the density and dynamic viscosity of the medium,
       respectively. $g$ is the acceleration due to gravity and $r$ is the 
radius of the colloids. For particles, we can approximate $\eta' \gg \eta$;
and for bubbles, $\eta'=0$ and $\rho'=0$. Hence, according to the 
above equation, the terminal velocity of colloids with size in nano range 
have a negligibly small terminal velocity. 
These colloids also experience a random force that originates from fast 
collisions with molecules of the medium. This gives rise to Brownian motion of
 the colloids. 
The average squared displacement of the colloid, $\langle x^2 \rangle $, that 
follows Brownian motion, 
after time $t$ from its initial position is given by (Reif 1965)
\begin{equation}
\langle x^2 \rangle=( \frac {k_B T}{3\pi \eta r}) t
\label{diffusion}
\end{equation}
 The average squared displacement is inversely proportional 
to the radius of the colloid and to the viscosity of the medium. Hence, 
diffusive motion dominates the motion of the particles and bubbles. 
These particles and bubbles encounter during their random motion 
inside the solution, and during this process, the particles get attached 
with the bubbles. It should be noted that particle-bubble attachment can 
occur when particle-bubble contact time is 
longer than the induction time (Dai et al. 1999). Hence, reduction in induction 
time 
enhance the attachment efficiency. As bubbles and particles are very small in 
size, the induction time is very small and attachment efficiency is very high.
 The  induction time is defined as the time for the liquid film between the 
particle and the bubble to thin, rupture and form a equilibrium three-phase 
contact.  

It is  observed that CdTe nanoparticles spontaneously aggregate
into a pearl-necklace like structure upon controlled removal of the protective
shell of organic stabilizer, and subsequently recrystallize into
nanowires (Tang et al. 2002). These CdTe nanoparticles have a large dipole 
moment 
and the dipole-dipole interaction between them is responsible for their 
unidirectional self-organization (Sinyagin et al. 2005). 
CdS nanoparticles with wurzite crystal structure also have a large dipole 
moment (Sinyagin et al. 2005; Blanton et al. 1997; Shanbhag and Kotov 2006).
 Hence, particle-attached-bubble as a 
whole may has a strong net dipole moment; and  dipole-dipole interaction 
between these particle-attached-bubbles is primarily responsible for their
 unidirectional aggregation.
The adjacent nanoparticles on the bubble surface probably get attached at a 
planar interface, reduce total surface energy and transform into a stable 
nanotube. In the aggregation-based crystal growth ( Banfield et al. 2000;
Penn and Banfield 1998), 
random force imparted on the nanoparticles by the medium molecules helps them 
in rotation and attachment at a planar interface so that they 
 can share a common crystallographic orientation. However, even a small 
misorientation can lead to dislocation at the interfaces. As the medium is 
viscous and CdS nanoparticles 
are attached on a curved surface with equilibrium three phase contact, 
particles could not perfectly orient and attach in a atomically flat 
interfaces to give a dislocation free single crystalline nanotube. However,
annealing after synthesis can help in improving the crystallinity.

\begin{figure}
\begin{center}
\epsfxsize=8.8true cm{\epsfbox{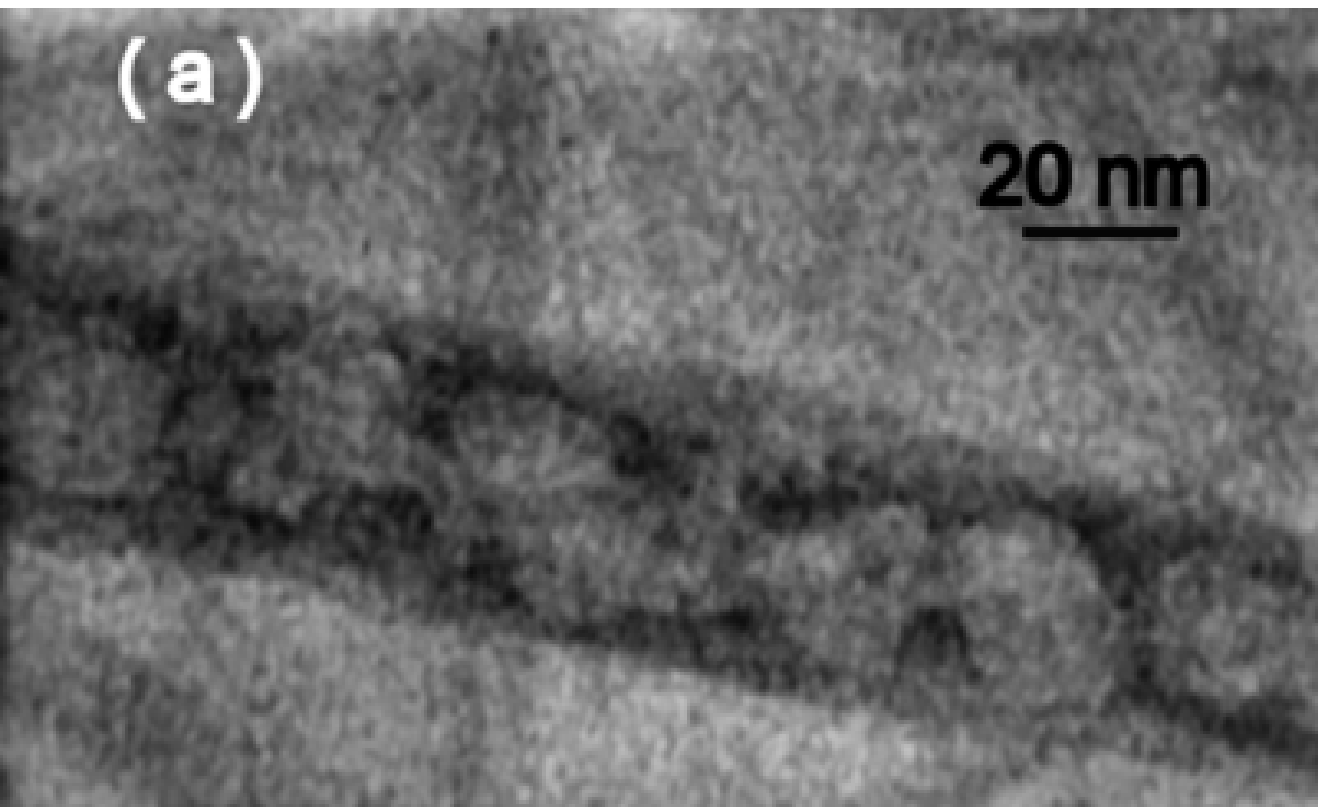}}
\epsfxsize=8.8true cm{\epsfbox{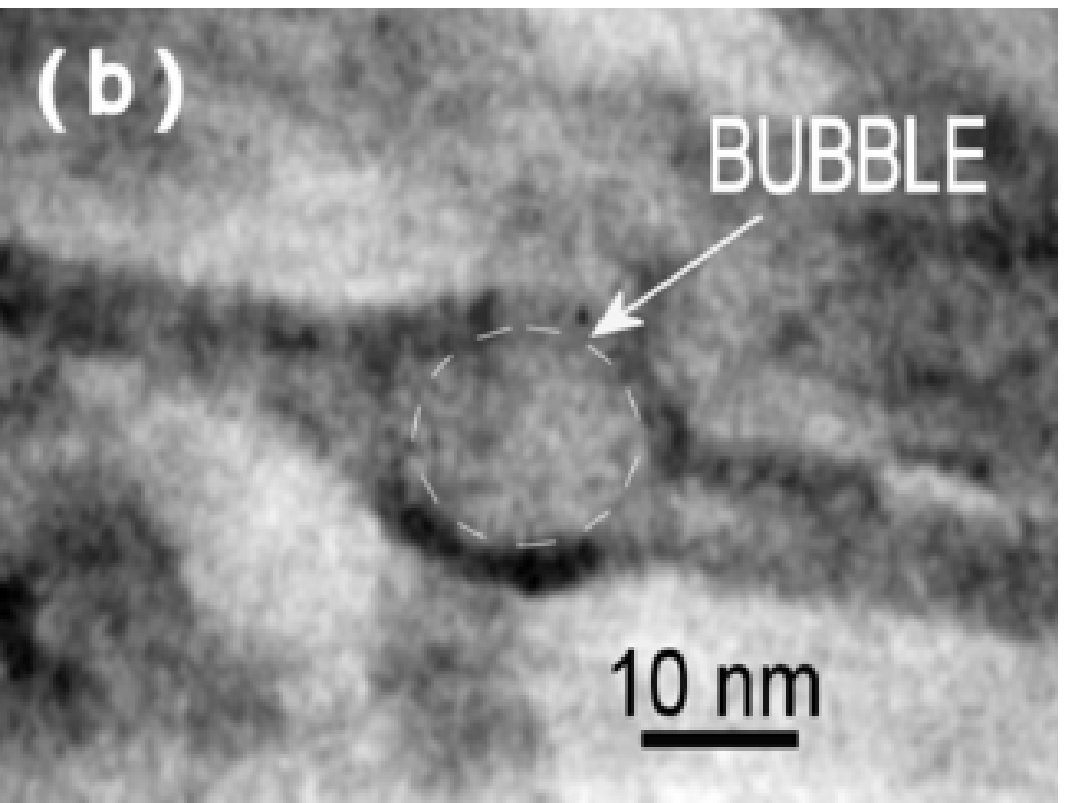}}
\caption{ TEM micrographs of (a) a nanotube in the  
formative stage (b) a nanowire with a rounded bulge. }
\label{bulge}
\end{center}
\end{figure}

It should be noted that colloids after nucleation goes through subsequent 
growth dynamics along with the above mention steps like bubble-particle 
attachment and unidirectional aggregation of particle-attached-bubbles. The 
colloids smaller than its critical size  dissolve as surface energy is large. 
Colloids bigger than the critical size only grow. The 
attachment of colloids also reduces the total surface energy and  effective 
size of the colloid become more than its critical radius. Hence, formation of
nanotubes with a high aspect ratio are favorable and stable. In the TEM 
measurement, it is observed that inner wall of nanotubes are comparatively 
less smooth than the outside wall. A  TEM micrograph of a nanotube in a  
formative stage, in which bubbles are coalescing, is shown in 
figure (Fig.\ref{bulge}a). Other possible nanostructures like nanoparticles, 
nanowires, nanowires with spherical cavity(Fig.\ref{bulge}b) are also 
observed in TEM measurements, although they are few in number. These 
observations lead to believe that bubbles are responsible for  
hollowness inside the nanotube. 

\begin{figure}
\begin{center}
\epsfxsize=8.8true cm{\epsfbox{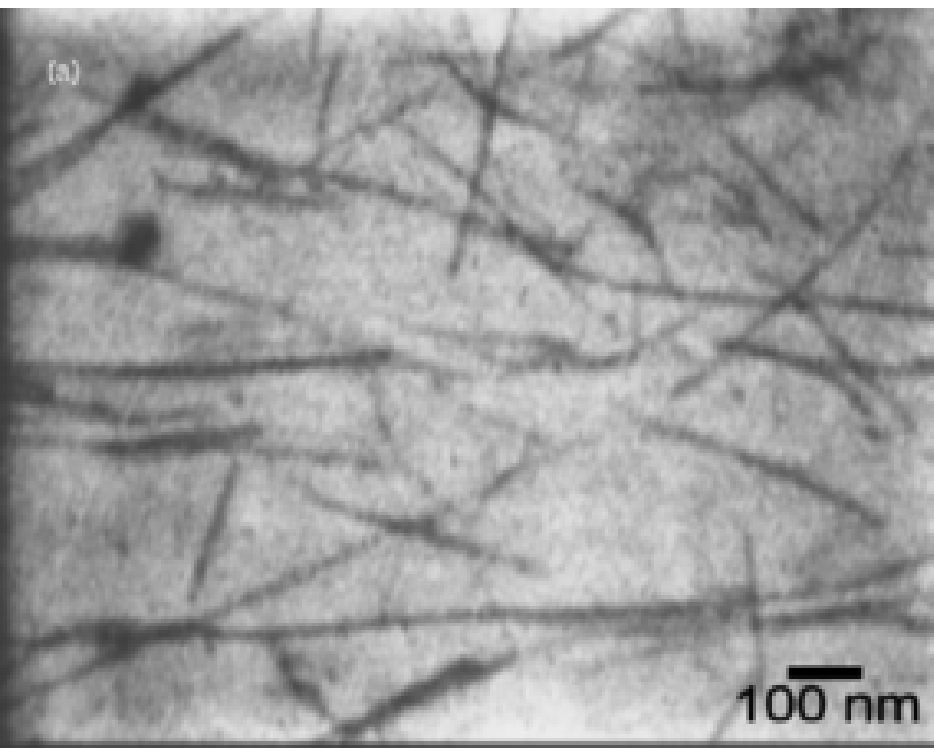}}
\epsfxsize=8.8true cm{\epsfbox{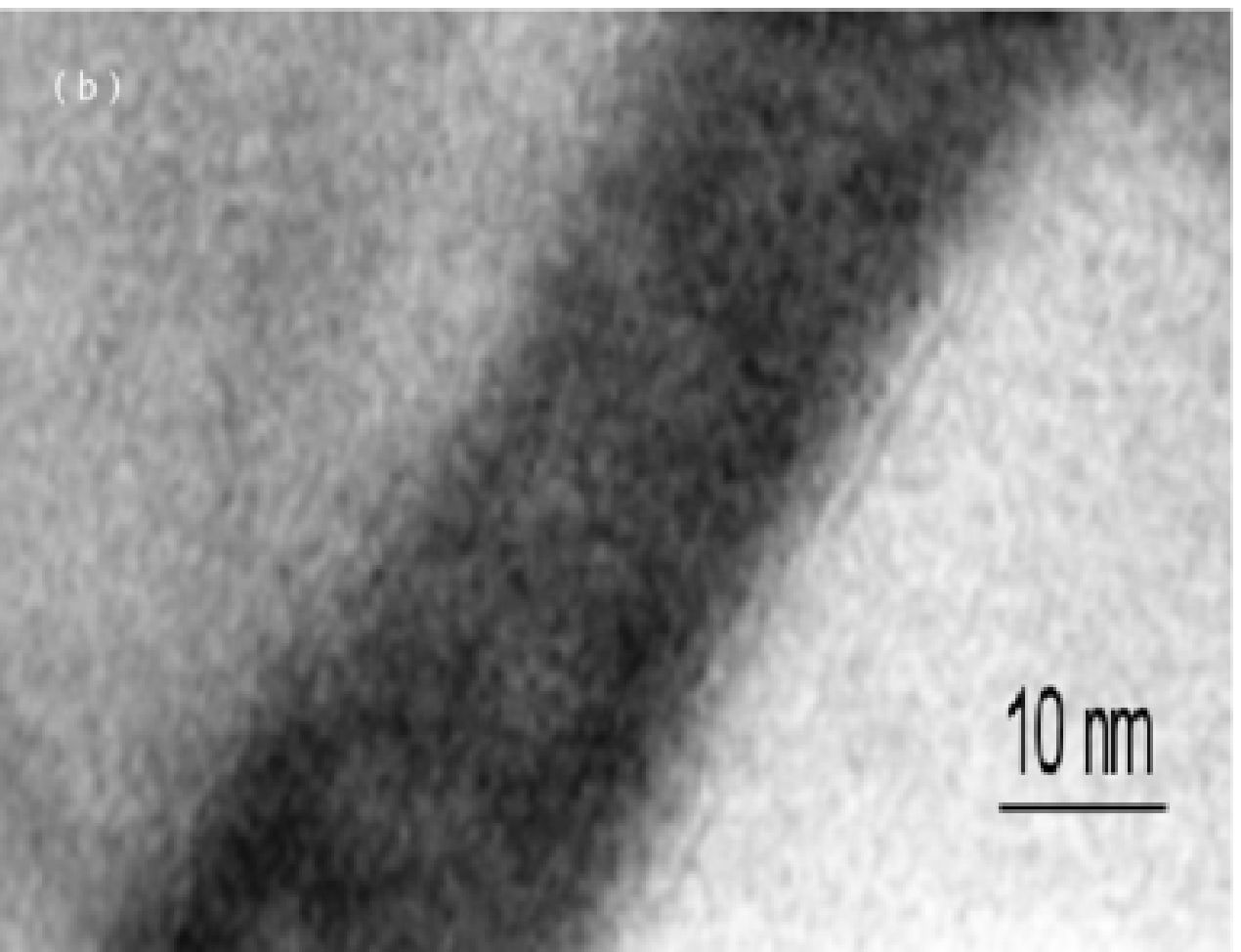}}
\caption{TEM micro-graphs of (a) CdS nanowire with lower magnification, and  
(b) a single nanowire with higher magnification.}
\label{wire}
\end{center}
\end{figure}

In order to confirm the role of PVA, experiments are carried out in  similar 
experimental condition by not using PVA, and by using much higher amount 
of PVA (25 mL of 20$\%$ aquous solution). Nanowires are seen in TEM measurements instead of 
nanotubes, when PVA is not used in the experiment. The typical TEM micrographs 
at a lower and at a higher magnification are shown in Fig.\ref{wire}a,b
 respectively.
In this case probably bubbles did not nucleate. Nanoparticles get aligned 
unidirectionally due to their dipole moment and nanowires are formed. 
Formation of nanowires even in the absence of PVA exclude the 
possibility that linear chain structure of polymer is someway acting as a 
template for unidirectional aggregation of the nanoparticles, and forming 
nanotubes. 

 \newpage

\begin{figure}
\begin{center}
\epsfxsize=8.8true cm{\epsfbox{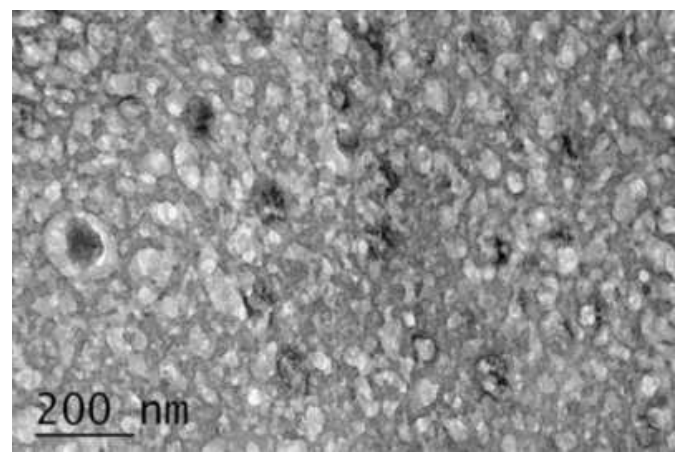}}
\caption{TEM micro-graph of CdS nanocrystals.}
\label{nanocrystal}
\end{center}
\end{figure}

When excess amount of PVA is used, the solution becomes very 
viscous. Only after $\approx$4 h ( in contrast to $\approx$10 min), 
the solution becomes pale yellow color. 
The solution then kept for another 90 min, and then spin-coated on a 
Cu-grid to carry out TEM measurements. Even though TEM is carried out by taking the grid on a liquid nitrogen cooled sample stage, the PVA matrix  forms voids 
due to the heating effect of electron beams. However, it can be seen that 
nanocrystals are well separated. With too much increase in viscosity, the 
decrease in diffusive length of nanocrystals is significant 
(see Eq. \ref{diffusion}). So nanocrystals could not aggregate to form 
nanowires or nanotubes. A typical TEM micrograph is shown in 
Fig. \ref{nanocrystal}.  It should be noted that well separated HgS 
nanocrystals are also synthesized  by a similar chemical procedure using 
PVA (Mahapatra and Dash 2006).  

  Most of the interesting properties exhibited by semiconducting nanomaterials 
are attributed to quantum confinement effect.  The 
electronic energy levels are strongly dependent on the size and also on the 
shape of the nanostructure (Kayanuma 1991). In the one-dimensional systems
charge carriers are confined in two dimensions and free in one dimension.
The spatial confinement of carriers leads to band gap widening and most 
directly realized by a high energy shift in optical absorption and 
photoluminescence peak. Quantum confinement effect for zero-dimensional system
 is studied  both theoretically (Kayanuma 1988) and 
experimentally (Vossmeyer et al. 1994) in detail. However, it is not well 
understood for one-dimensional systems, particularly with tubular structure.  
 
\begin{figure}
\begin{center}
\epsfxsize=8.8true cm{\epsfbox{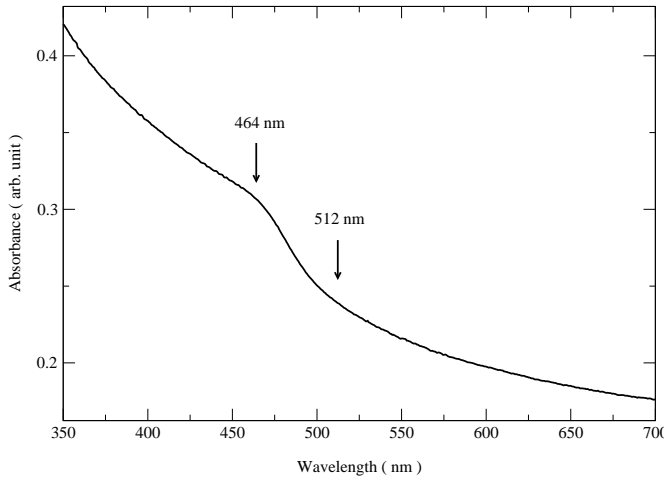}}
\caption{Optical absorption spectrum of CdS nanotubes.}
\label{optical}
\end{center}
\end{figure}

\begin{figure}
\begin{center}
\epsfxsize=8.8true cm{\epsfbox{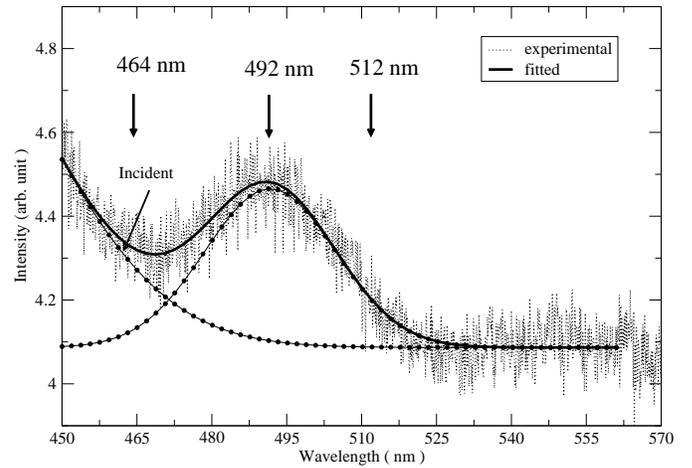}}
\caption{Photoluminescence spectrum of CdS nanotubes. Experimental data is 
fitted with Gaussian functions.}
\label{pl}
\end{center}
\end{figure}

 The optical absorption spectrum and photoluminescence (PL) spectrum of  
CdS nanotubes are shown in Figs. \ref{optical}, \ref{pl}, respectively.
 An excitonic peak appears at 
464 nm in the optical absorption spectrum. The peak position is blue shifted 
by 48nm  from  its bulk band gap value (512 nm). The experimental PL data is 
fitted as the sum of two Gaussian functions, among which one is due to incident line. The PL peak is
 best fitted with the Gaussian of 492 nm
mean and 26 nm standard deviation. It should be noted that,
the PL peak position is also blue shifted from its bulk band gap value.
High energy peak shifting in the optical absorption and PL spectra are expected
due to quantum confinement effect as nanotube wall thickness is comparable to 
excitonic diameter of bulk CdS.

 Luminescence that are observed in semiconductor nanomaterials are generally 
excitonic and trapped emissions. Excitonic 
emission is sharp and can be observed near the absorption edge if the material 
is pure (Pankove 1975). However, if the material is impure or 
off-stoichiometric 
then a broad 
and intense emission occur at higher wavelength due to recombination of 
charge carriers at trapped states. Hence the band edge luminescence that 
appears at 492 nm is due to  excitonic transition. Band edge luminescence at 
470 nm is also observed for CdS nanotubes synthesized by sacrificial template 
method (Li et al. 2006). However, CdS nanotubes synthesized by sacrificial 
template method also show a very intense and broad peak centered around 560 nm 
due to presence of trapped states. No such peak is observed in our synthesized 
CdS nanotubes. This suggests purity and stoichiometric nature of our 
synthesized CdS nanotubes.    

   Excitonic transition in a semiconductor can be observed well only at low 
temperature. However, it can be observed even at room temperature due to 
enhancement of oscillator strength in the low-dimensional systems 
(Kayanuma 1991). It should be noted that excitonic 
transitions is not observed in optical absorption and PL measurements for bulk
 CdS which was prepared by the similar chemical route (Pavaskar et al. 1977). 
However, a clear excitonic feature arises in both absorption and PL measurements
when it forms nanotubes. The PL peak is red shifted by 150 meV with respect to absorption
 peak. Such a large red shift, known as stokes shift, of 147 meV is also 
reported for CdS nanoparticles (Tamborra et al. 2004).
 The difference in the absorption and emission states avoid sample 
self-absorption and could be very useful in making LEDs (Sze 1981).

\section*{Conclusion}
 CdS nanotubes with wall thickness comparable to excitonic diameter 
of the bulk material are synthesized by a chemical synthesis process. These 
synthesized  nanotubes show band gap widening and enhanced oscillator strength
due to quantum confinement effects. A large stokes shift of 150 meV is also 
observed. Bubbles are  responsible for the hollowness of  
nanotubes; and bubbles of dissolved gases can be utilized to make  
nanostructures with hollow interior.

\section*{Acknowledgment}
 The help and encouragement received from Dr S. N. Sahu is gratefully 
acknowledged. Mr A. K. Dash, Mr U. M. Bhatta, and Dr P. V. Satyam are 
acknowledged for their help in TEM measurements.


\begin{thebibliography}{99}
\bibitem {banfield}Banfield JF,  Welch SA, Zhang H, Ebert TT, Penn RL (2000)
Aggregation-based crystal growth and microstructure development in natural 
iron oxyhydroxide biomineralization products. Science 289:751-754 
\bibitem {bhattacharya} Bhattacharya A, Ray  P (2004) Studies on surface 
tension of poly(vinyl alcohol): effect of concentration, temperature, and 
addition of chaotropic agents. J Appl Polym Sci 93:122-130 
\bibitem{blanton}  Blanton SA, Leheny RL, Hines MA, Sionnest PG (1997) 
Dielectric dispersion measurements of CdSe nanocrystal colloids: Observation 
of a permanent dipole moment. Phys Rev Lett 79:865-868 
\bibitem{bowers}Bowers PG, Hofstetter C, Letter  CR, Toomey RT (1995) 
Supersaturation limit for homogeneous nucleation of oxygen bubbles in water at elevated pressure:``superhenry's law". J Phys Chem 99:9632-9637 
\bibitem {briscoe}Briscoe B, Luckham P, Zhu S (2000) The effects of hydrogen
bonding upon the viscosity of aqueous poly(vinyl alcohol) solution. 
Polymer  41:3851-3860 
\bibitem {chopra}Chopra NG, Luyken RJ, Cherrey K, Crespi VH, Cohen ML, 
Louie SG, Zettl  A (1995) Boron nitride nanotubes. Science 269:966-967 
\bibitem{dai} Dai Z, Fornasiero D, Ralston J (1999) Particle-bubble attachment 
in mineral flotation. J Colloid Interface Sci 217:70-76
\bibitem {lieber} Duan X, Lieber CM (2000) General synthesis of compound 
semiconductor nanowires. Adv Mater 12:298-302 
\bibitem{duan}Duan X, Wang J, Lieber CM (2000) Synthesis and optical 
properties of gallium arsenide nanowires. Appl Phys Lett 76: 1116-1118 
\bibitem{feldman} Feldman Y, Wasserman E, Srolovitz  DJ, Tenne  R (1995) 
High-rate, gas-phase growth of $MoS_2$ nested inorganic fullerenes 
and nanotubes. Science 267: 222-225 
\bibitem{hariskos} Hariskos D, Powalla M, Chevaldonnet N, Lincot D, 
Schindler A, Dimmler  B (2001) Chemical bath deposition of CdS buffer layer:
prospects of increasing materials yield and reducing waste. Thin Solid Films 
387: 179-181 
\bibitem {hu}Hu  JQ, Bando Y, Zhan JH, Liao MY, Golberg D, Yuan XL,
 Sekinuchi T (2005) Single-crystalline nanotubes of IIB-VI semiconductors.
 Appl Phys Lett 87: 113107 
\bibitem {iijima}Iijima (1991) Helical microtubules of graphitic carbon.
 Nature 354: 56-58 
\bibitem {kayanuma} Kayanuma Y (1988) Quantum-size effects of interacting 
electrons and holes in semiconductor microcrystals with spherical shape. 
Phys Rev B 38: 9797-9805 
\bibitem{kaya} Kayanuma Y (1991) Wannier excitons in low-dimensional 
microstructures: shape dependence of quantum size effect. 
Phys Rev B 44: 13085-13088 
\bibitem {lamb} Lamb  H (1945) Hydrodynamics, 6th edn. Dover,  New York
\bibitem {landau} Landau LD, Lifshitz EM (1999)
 Statistical physics, 3rd edn. Butterwort Heinemann 
\bibitem{li} Li X , Chu H , Li Y (2006) Sacrificial template growth of CdS 
nanotubes from $Cd(OH)_2$ nanowires. J Solid State Chem 179:96-102 
\bibitem{mahapatra} Mahapatra AK, Dash AK (2006) $\alpha$-HgS 
nanocrystals: synthesis, structure and optical properties. Physica E 35: 9-15
\bibitem{markov} Markov IV (1995) Crystal growth for beginners, World scientific
\bibitem{masale} Masale M, Constantinou NC, Tilley DR (1992) Single-electron 
energy subbands of a hollow cylinder in an axial magnetic field.
 Phys Rev B 46: 15432-15437 
\bibitem {pankov} Pankove JI (1975) Optical process in semiconductors, Dover 
Publications Inc., N.Y.
\bibitem {pavaskar} Pavaskar NR, Menezes CA, Sinha APB (1977) 
Photoconductive CdS films by a chemical bath deposition process. J Electrochem 
Soc 124: 743-748 
\bibitem {penn} Penn  RL, Banfield JF (1998) Imperfect oriented attachment: 
dislocation generation in defect-free nanocrystals. Science 281: 969-971 
\bibitem {reif} Reif F (1965) Fundamentals of statistical  and thermal physics,
 McGraw-Hill, New York 
\bibitem {sachin} Shanbhag S, Kotov NA (2006) On the origin of a permanent 
dipole moment in nanocrystals with a cubic crystal lattice: effects of 
truncation, stabilizers, and medium for CdS tetrahedral homologues. 
J Phys Chem B 110:12211-12217 
\bibitem {sinyagin} Sinyagin A, Belov A, Kotov N (2005) Monte Carlo simulation 
of linear aggregate formation from CdTe nanoparticles. Model Simul Mater 
Sci Eng 13:389-399
\bibitem {sze} Sze SM (1981) Physics of semiconductor devices, 
Wiley Interscience 
\bibitem{tamborra} Tamborra M, Striccoli M, Comparelli R, Curri ML, 
Petrella A, Agostiano A (2004) Optical properties of hybrid composites
based on highly luminescent CdS nanocrystals in polymer.
Nanotechnology 15:S240-S244  
\bibitem {tang} Tang  Z, Kotov NA, Giersig M (2002) Spontaneous organization 
of single CdTe nanoparticles into luminescent nanowires. Science  297:237-240 
\bibitem{tenne} Tenne R, Margulis L, Genut M, Hodes G (1992) Polyhedral and 
cylindrical structures of tungsten disulfhide. Nature 360:444-446 
\bibitem{vossmeyer} Vossmeyer T, Katshikas L, Giersig M,  popovic IG, 
Weller H (1994) CdS nanoclusters: synthesis, characterization, size dependent 
oscillator strength, temperature shift of the excitonic transition energy, and 
reversible absorbance shift. J Phys Chem 98:7665-7673 
\bibitem{yu} Yu DP, Bai ZG, Ding Y, Hang QL, Zhang HZ, Wang JJ,
Zou YH, Qian W, Xiong GC, Zhou HT, Feng SQ (1998) Nanoscale silicon wires 
synthesized using simple physical evaporation. Appl Phys Lett 72:3458-3460
\bibitem {zhan} Zhan J, Yang X, Wang D, Li S, Xie Y, Xia  Y, Qian Y (2000) 
Polymer-controlled growth of CdS nanowires. Adv Mater 12:1348-1351
\end{thebibliography}
\end{document}